\documentclass[%
 aip,
rsi,%
 amsmath,amssymb,
 reprint,%
]{revtex4-1}

\usepackage{graphicx}
\usepackage{dcolumn}
\usepackage{bm}

\begin{document}

\preprint{AIP/123-QED}

\title[Alignment of a Vector Magnetometer to an Optical Prism]{Alignment of a Vector Magnetometer to an Optical Prism}

\author{M. R. Dietrich}
\email{mdietrich@anl.gov}
\author{K. G. Bailey}%
\author{T. P. O'Connor}
\affiliation{ 
Physics Division, Argonne National Laboratory, Lemont IL 60439, USA
}%

\date{\today}

\begin{abstract}
A method for alignment of a vector magnetometer to a rigidly attached prism is presented.  This enables optical comparison of the magnetometer axes to physical surfaces in an apparatus, and thus an absolute determination of the magnetic field direction in space.  This is in contrast with more common techniques, which focus on precise determination of the relative angles between magnetometer axes, and so are more suited to measuring differences in the direction of magnetic fields.  Here we demonstrate precision better than 500 $\mu$rad on a fluxgate magnetometer, which also gives the coil orthogonality errors to a similar precision.  The relative sensitivity of the 3 axes is also determined, with precision of about 5$\times 10^{-4}$.
\end{abstract}

\maketitle

\section{\label{sec:Introduction}Introduction}

To use a vector magnetometer for precision measurement of magnetic field direction, it is necessary to calibrate not only the relative sensitivity of its three axes, but also to determine any orthogonality error between those axes.  Typically, this is done by rotating the magnetometer in a controlled way in a static magnetic field, and then fitting the resulting data points to a model that includes a list of possible systematic errors\cite{Pang2013}.  This results in a highly accurate measurement of the relative magnetometer axes, but those axes are oriented somewhat arbitrarily in space.  This is sufficient to measure changes in the magnetic field direction and amplitude, and so the resulting calibration can be used to determine spatial gradients, or to measure changes in time, such as during attitude adjustment in a satellite.  It is sometimes necessary, however, to determine the absolute direction of the magnetic field in some location compared to nearby physical surfaces, and thus a direct calibration against some external coordinate system is desired.  For our application, we desired to guarantee alignment of magnetic fields to an electrode surface inside a nearby vacuum system at a level better than 1 mrad.

To address this problem, we chose to rigidly attach an optical alignment cube to a fluxgate vector magnetometer.  By rotating this structure inside of a 3D Helmholtz coil, which was driven with an AC signal, we were able to align the three fluxgate axes to the coordinate system defined by the alignment cube.  This entailed a least squares fit to the coil fields as well as the magnetometer axes.  Although similar techniques have been proposed before,\cite{McPherron1978} we here use the method to align the magnetometer permanently to an external optical reference.  One strength of this approach is its simplicity: because the coil axes are simultaneously determined by the fit, it is not necessary to use a set of coils which are highly orthogonal to one another.  Ambient fields are mitigated due to the lock-in measurement, and precise rotations of the magnetometer structure is provided by the prism itself, so a non-magnetic rotation mount or theodolite is unnecessary.

\section{\label{sec:Setup}Experimental Setup}

The basic principle of the calibration is to use a set of Helmholtz coils to produce 3 nearly orthogonal fields, and to measure each of the field directions with the vector magnetometer.  From these three measurements alone, we can not determine what part of the result is due to orthogonality error in the fluxgate, orthogonality error in the coils, or simple rotation between the fluxgate and the coils.  However, if we rotate the fluxgate into at least 3 perpendicular positions, making sure to permute every axis, and repeat the three measurements in each position, then the contribution from each of those errors can be independently ascertained.  By solving the relevant system of equations one obtains the coil and fluxgate axes in the (unique) lab-fixed frame that is described by that set of rotations.  By precisely aligning this coordinate system with a prism rigidly attached to the magnetometer, we can evaluate the magnetometer axes in terms of the prism axes, and by extension, in terms of one another.

\begin{figure}
\includegraphics[width=\linewidth]{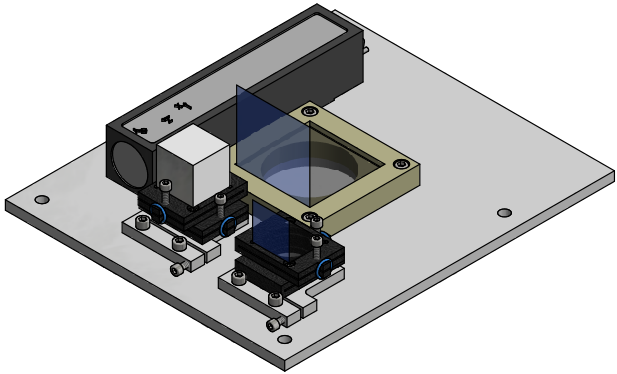}
\caption{\label{fig:fgassembly}A drawing of the fluxgate assembly, with the alignment cube in white, and the right angle prisms in translucent blue.  The plate is made from macor.  The alignment cube sits on a plastic mount so it can be aligned with the right angle prism.  Notice that neither right angle prism is used in the calibration, they are for use after the fluxgate is installed in its final application.  For reference, the alignment cube is 1 inch on a side.}
\end{figure}

A fluxgate vector magnetometer and an alignment cube were attached by nylon screws and epoxy to a Macor plate, custom machined to fit in its intended application, see figure \ref{fig:fgassembly}.  These materials were chosen to avoid any metal, since even a non-magnetic conductor would shield the AC fields we use in calibration.  Two such plates were created, as per our experimental requirements, although this also provides for a cross-calibration, discussed below.  The alignment cube is specified so that each edge has an angle error of less than 15 $\mu$rad, and five of its surfaces are coated with aluminum mirror surfaces.  When put in place, the prism was also mechanically constrained to align with the nominal axes of the fluxgate, to ensure that the misalignment errors are small, as assumed in the analysis below.  We also attached right angle prisms to the same Macor plate, so that the coordinate prism could be compared to objects below the Macor plate.  To complete the calibration, we need to not only rotate the Macor plate in its own plane, but also tilt it upwards by 90 degrees, so as to permute the vertical axis.  A separate plastic stand was produced to enable this rotation, which was designed to maintain the prism at a constant height.

\begin{figure}
\includegraphics[width=\linewidth]{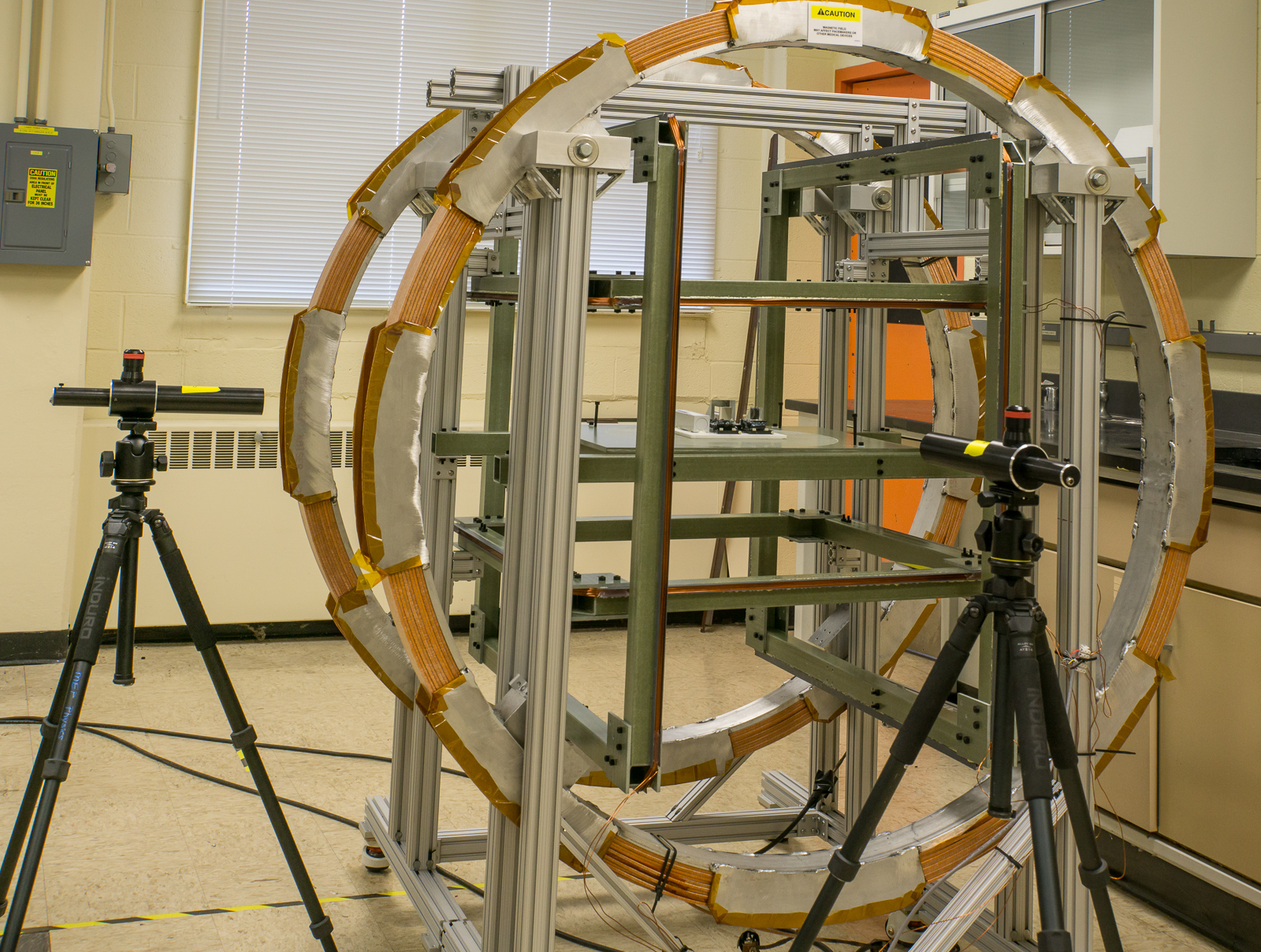}
\caption{\label{fig:coils}A picture of the calibration setup.  The autocollimators can be seen to the left and right, on tripods.  The macor plate is on the garolite platform in the coil assembly.  The inner coils are rectangular and green, and the outer coil is circular.  The alignment cube can be seen in the very center.}
\end{figure}

A set of 3 orthogonal Helmholtz coils were obtained, see figure \ref{fig:coils}.  The superstructure of this coil was made from aluminum and stainless steel fittings, but the coil forms are plastic (G10 garolite).  To eliminate the effect of fluxgate offsets and ambient magnetic fields, measurements were performed with AC magnetic fields and software lock-in detection.  Although the aluminum superstructure can attenuate the AC fields, no shielding of any coil was detected below 5 Hz, and so the calibration was performed at 2 Hz.  The sign of the magnetic field is important, and is determined by rounding the cosine of the phase to the nearest integer.  The field amplitude was approximately 10 $\mu$T.  A garolite platform was attached to the aluminum frame to support the magnetometer, and plastic trim screws installed in this platform to allow careful alignment of the Macor plate to the autocollimators after each rotation.  This proved essential, since small flexures in the plastic table caused imperfect rotations, clearly visible on the autocollimators, which needed to be corrected.  We find generally that our calibration was limited by the precision of this trimming, and future iterations should include a more elaborate rig that allows fine tuning in the horizontal and vertical angles.

Any environmental background at the lock-in frequency would present a systematic.  Such a bias can be detected as a non-null signal in the magnetometer when the coils are shorted, and mitigated by subtracting this background from each measurement, or potentially, by choosing a new frequency.  If a background is subtracted from each measurement, it should be independently measured in each position, to account for rotations and gradients.  However, no such AC bias was detected in our apparatus at the 10 part-per-million level.

The fluxgate is initially aligned to the coils by simply applying a current to one coil and adjusting the fluxgate position manually until one detector axis is maximized.  Then, two autocollimators, mounted on tripods, are placed on the floor a meter or two from the fluxgate, and aligned carefully to retroreflect off the prism.  These two autocollimators are now perpendicular, by virtue of the prism's squareness, and define the lab coordinate system discussed below.  Since the fluxgate has already been aligned to a coil axis, the lab coordinate system is also roughly aligned to the coils.  The autocollimators do not move for the remainder of the calibration.  To perform a rotation, one simply rotates the macor plate, finely adjusting its angle until retroreflection is achieved simultaneously on \emph{both} autocollimators.

It is essential that the prism and magnetometer be placed as close to one another as possible.  As the rotations are performed, it is the prism that must stay in place, and the magnetometer that moves, since the prism must stay within view of the autocollimators at all times.  Thus, any gradients introduced by the finite size of the coils will be sampled as the rotations are performed, based on this distance.  Roughly, the radius of the coils should be at least 7 times the distance between the prism and magnetometer to achieve a precision of .1\% \cite{Cacak1969,Nissen1996}.  Due to the geometric constraints of our apparatus, the prism is in one case 5 cm from the magnetometer, and so our smallest coil is a square 70 cm on a side.

\section{\label{sec:Analysis}Analysis}

There are three relevant coordinate systems; the lab coordinate system, described by the two autocollimators, then the prism coordinate system, and finally the fluxgate coordinate system.  We will assume that the lab and prism coordinate systems are \emph{approximately} aligned to the fluxgate, at the level of about 1 degree (17 mrad) so that we can neglect their misalignment in second order.  However, alignments between the prism and lab frame can be done with extremely high precision, due to the autocollimators, at a level better than 100 $\mu$rad, and therefore much more accurately than the target sensitivity of this calibration.  The prism coordinate system is related to the fluxgate coordinate system by a fixed transformation matrix we will call $\mu$, whose determination is the primary objective of this analysis.  We will label the rotations with the index $i$.  The unitary matrix describing the rotation between the prism coordinate system and the lab system is $R_i$.  For each rotation, we create and measure 3 roughly orthogonal fields.  Each of these fields will be one column of the $3\times 3$ matrix $G$, which is the magnetic field amplitude as represented in the lab frame.  Similarly, the result of each measurement will be one column of $F_i$, which is represented in the magnetometers's coordinate system.  Finally, we will define a matrix $S$, which is diagonal and describes the relative sensitivity of each magnetometer axis.  In terms of these elements, the magnetic field, as described in the prism frame, is given by $B_P = R_i G$.  The magnetic field in the prism frame can also be determined by applying the (unknown) correction matricies to the magnetometer's measurement, which can be written $B_P = \mu S F_i$.  Thus, we have the matrix equation

\begin{equation}
\label{eq:master}
\mu S F_i = R_i G.
\end{equation}

By our representation of $R_i$, we choose the coordinate system for the lab and $G$.  That is, it is the choice of rotations itself which defines the lab coordinate system, and we must make this correspond to the autocollimators by performing all rotations accurately around those instruments.  That is how this technique connects the magnetometer frame to the prism frame, even though there is no direct measurement sensitive to both.

The expression \ref{eq:master} has 18 unknowns (6 from $\mu$, 9 from $G$, and 3 from $S$) and each rotation provides 9 equations, so it might seem only 2 rotations are required.  However, those two positions will not uniquely define the lab coordinate system, and so the operation gives inaccurate results without at least 3.  The problem is then overconstrained, and so we seek a best fit solution.  The corresponding least-squares problem gives a trivial solution, unless appropriate constraints are applied to the unknowns.  We will rewrite equation \ref{eq:master} to include these (nonlinear) constraints implicitly, and linearize the result to guarantee a unique solution.

First, we would like to normalize the magnetic fields to be nearly one by multiplying equation \ref{eq:master} on the right by $N$.  This matrix is nominally arbitrary, but we will find it convenient to choose a diagonal matrix where, for instance, $N_{xx} = 1/\overline{F_{xx}}$, and $\overline{F_{xx}}$ is the average field strength of the $X$ Helmholtz coil as measured by each of the 3 magnetometer axes.  With this choice, the matrices $R_i^T F_iN$ and $GN$ will both be approximately equal to the identity matrix, and so we can write

\begin{equation}
\begin{aligned}
\mu S R_i R^T_i F_i N & = R_i G N\\
\mu S R_i (I + F_i^\Delta) & = R_i (I + G^\Delta)
\end{aligned}
\end{equation}

\noindent where $F_i^\Delta$ and $G^\Delta$ are small.  Similarly, we can write $\mu = I + \mu^\Delta$ and $S = I + S^\Delta$.  After removing products that are second order in smallness and collecting terms we arrive at

\begin{equation}
\label{eq:solve}
(\mu^\Delta + S^\Delta)R_i - R_i G^\Delta = R_i - F_i N 
\end{equation}

\noindent Since $\mu^\Delta$ has only off-diagonal elements at first order, and $S^\Delta$ has only diagonal elements, these matrices combine to form a single, easily separated matrix $U^\Delta$ which encapsulates all the magnetometer errors.

To solve the least-squares problem that results from the several rotations, we would like to write equation \ref{eq:solve} in the form $Ax = b$, but at first glance it has entirely the wrong structure.  We need to vectorize\cite{Laub} the 18 unknowns from $U^\Delta$ and $G^\Delta$ into a single column vector $x$.  The 9 known parameters from the right hand side are flattened in the same, row-major, way to a 9 element column vector $b_i$.  If the system of equations \ref{eq:solve} is rewritten in this way, the $9 \times 18$ coefficient matrix $A_i$ is given by 

\begin{equation}
A_i = \left( I_3 \bigotimes R^T_i, -R_i \bigotimes I_3 \right),
\end{equation}

\noindent where $\bigotimes$ is the Kronecker product.  By concatenating the several $A_i$ and $b_i$, one obtains a final, overdetermined matrix equation in the form $Ax = b$.  If there are $M$ rotations, then the matrix $A$ has dimensions $9M\times 18$, the matrix $x$ is $18\times 1$ and $b$ is $9M\times 1$.  The first nine elements of $x$ correspond to $U^\Delta$, and the second nine to $G^\Delta$.  Once in this form, the least-squares problem can be solved using one of many standard techniques, such as the singular value decomposition\cite{Laub}.  Armed with this solution, we can form the correction matrix $\mu S \approx (I+\mu^\Delta )(I+S^\Delta ) \approx I + U^\Delta$, which is used to accurately rotate any magnetic field measurement into the prism's coordinate frame.

Equation \ref{eq:master} can also be solved directly, by minimizing

\begin{equation}
\left| \mu S F_i - R_i G \right|^2
\end{equation}

\noindent subject to the 7 nonlinear constraints

\begin{equation}
\begin{aligned}
\rm{diag} \left(\mu^T \mu \right) & = 1 \\
\rm{diag} \left(N^T G^T G N\right) & = 1 \\
\rm{tr} (S) &  = 3
\end{aligned}
\end{equation}

\noindent which can be introduced as Lagrange multipliers.  This can allow relaxation of the requirement that the squares of the alignment errors be negligible, and therefore achieve greater precision, but care must be taken to ensure the global minimum is found.  This can be facilitated by comparing with the solution to equation \ref{eq:solve}.

\section{\label{sec:Results}Results}

Two fluxgates were thusly calibrated using the same lab coordinate system; that is, the autocollimators and Helmholtz coils were not moved in between calibrations.  The fit procedure described above was used to obtain $\mu$, $S$ and $G$ for each fluxgate setup.  The uncertainty was estimated by calculating the product $R_i^T \mu S F_i$ for each measured $F_i$.  Since it is equal to $G$, this product should be the same for every rotation.  We found for the first fluxgate an average standard deviation of $5\times 10^{-4}$ rad in each matrix element, and for the second an average standard deviation of $3\times 10^{-4}$ rad, with no single matrix element standing out as an outlier.  This uncertainty is consistent with our expectation based on the sensitivity of the autocollimator and our estimate of scatter in alignment angle during each rotation.  Five rotations were performed during data taking, and in analysis we excluded one or two at a time to test for sensitivity to the number of rotations, but the solution was robust against these tests, within the uncertainty of the measurement.  Orthogonality errors of the fluxgate can be obtained from the off-diagonal elements of the matrix product $\mu^T \mu$, which were found to be consistent with the factory specification for the fluxgate, as were the errors in the sensitivity matrix $S$.  Our choice for the normalization matrix $N$ was also found to be insensitive; different choices for the diagonal elements would change the mean of the sensitivity errors, but not change the difference of those errors, as expected, since we are sensitive only to relative sensitivities.  Finally, since the coils were unchanged between the measurements, we expect to obtain the same matrix $G$ for each calibration, which was indeed the case.  This provides a powerful validation of our technique.

\section{\label{sec:Conclusion}Conclusion}

We have presented a simple method for absolute calibration of a vector magnetometer to a optical prism reference, using only a 3D Helmholtz coil and two autocollimators.  The precision generally was 300-500 parts per million, although this appears to have been limited by our ability to rotate the magnetometer by hand precisely, and improved precision should be achievable with a better, non-metallic rotation mount, which we anticipate would reduce the errors below 100 parts per million.  The prism should also be very close to the magnetometer to enable this.  The presented technique also extracts the coil orthogonality errors and relative magnetometer sensitivities at a similar accuracy.

\begin{acknowledgements}
This work is supported by U. S. Department of Energy (DOE), Office of Science, Office of Nuclear Physics, under contract No. DE-AC02-06CH11357.
\end{acknowledgements}

\bibliography{aipsamp}

\end{document}